\documentclass[twocolumn,showpacs,preprintnumbers,amsmath,amssymb]{revtex4}

\usepackage{graphicx}
\usepackage{epsf}
\usepackage{graphicx}
\usepackage{amsmath}
\usepackage{amssymb}
\usepackage{subfigure}

\newcommand{\order  }{{\cal O}}

\newcommand{\Bra    }{\left\langle}
\newcommand{\Ket    }{\right\rangle}
\newcommand{\atanh  }{{\rm{atanh}}}

\newcommand{\bh     }{\mbox{\boldmath$h$}}

\newcommand{\bJ     }{\mbox{\boldmath$J$}}
\newcommand{\bc     }{\mbox{\boldmath$c$}}

\newcommand{\bsigma }{\mbox{\boldmath$\sigma$}}

\newcommand{\btau   }{\mbox{\boldmath$\tau$}}

\newcommand{\bpsi   }{\mbox{\boldmath$\psi$}}

\begin{document}
\title{Small-world hypergraphs on a bond-disordered Bethe lattice}
\author{D. Boll\'e, R. Heylen}
\affiliation{Katholieke Universiteit Leuven, Instituut voor
Theoretische Fysica, Celestijnenlaan 200D, B-3001 Leuven, Belgium}
\email{{desire.bolle,rob.heylen}@fys.kuleuven.be}

\begin{abstract}
We study the thermodynamic properties of spin systems with bond-disorder on small-world hypergraphs, obtained by superimposing a one-dimensional Ising chain onto a random Bethe graph with $p$-spin interactions. Using transfer-matrix techniques, we derive fixed-point equations describing the relevant order parameters and the free energy, both in the replica symmetric and one step replica symmetry breaking approximation. We determine the static and dynamic ferromagnetic transition and the spinglass transition within replica symmetry for all temperatures, and demonstrate corrections to these results when one step replica symmetry breaking is taken into account. The results obtained are in agreement with Monte-Carlo simulations.
\end{abstract}
\maketitle

\section{Introduction}
Many real-world systems and applications under scrutiny these days require random diluted graphs to model the interactions. These graphs often are of a bipartite nature, separating the interactions from the variables in the model, and hence allowing for more complex schemes of interactions. A simple, and much used, example of this type of graphs is the finitely connected random graph with $p$-spin Ising-like interactions. Such graphs are succesfully employed in the description and analysis of error correcting codes (e.g. the low density parity check codes, see \cite{gallager} - \cite{sourlas}), satisfiability problems (e.g. the k-sat problem, \cite{ksat}), modelling of structural glasses and spinglasses \cite{structglass}, proteomic networks \cite{proteomic}, etc. 
\\
Some of these systems, however, require not only long-range interactions to effectively describe the system, but moreover possess local structures. There is a notion of neighborhood or distance between spins, and spins that are close have additional interactions of another type than the long-range ones. It is in this context that small-world networks have been developed, to provide a simple model that combines both long-range interactions with local interactions. One such small-world network to which a lot of work has already been devoted (see \cite{reptrans} and references therein) is the random graph supplemented with a one-dimensional chain connecting all spins together, yielding the most simple notion of neighborhood: nearest-neighbor interactions. \\
In this work we study the thermodynamic properties of such a small-world graph, composed of a finite and constant number of $p$-spin interactions (a $p$-spin Bethe-lattice) and a one-dimensional chain interconnecting all spins. We calculate the self-consistent equations governing the statics of the system by employing transfer-matrix techniques, and solve these equations in the replica symmetric (RS) approximation to determine the spinglass transition and the static and dynamic ferromagnetic transitions as function of the system parameters. We focus especially on the influence of bond-disorder. The corresponding phase diagrams show reentrance effects, indicating  replica symmetry breaking (RSB), and we apply one-step replica symmetry breaking (1RSB) in the reentrance regions. The results obtained are compared with Monte-Carlo simulations. This paper extends our earlier work (see \cite{hypergraphs}), where we used analogous methods to study a small-world graph with poisson-distributed interactions. However, in that paper we did not consider bond-disorder and only discussed  the RS ferromagnetic results. \\
This paper is organized as follows. In Sec. \ref{model} we define the small-world model. In Sec. \ref{saddlesec} we derive the saddle-point equations, and apply  transfer matrix techniques to obtain the eigenvector equations determining the order parameter functions. In Sec. \ref{rs} we solve these eigenvector equations in the RS approximation, leading to self-consistent equations for the order parameter functions. Sec. \ref{1rsb} discusses the analog 1RSB equations. The RS and 1RSB results are presented in Sec. \ref{rsresults} and Sec. \ref{1rsbresults}. Finally, Sec. \ref{discussion} presents some concluding remarks.

\section{The model} \label{model}
Consider a spin system of $N$ Ising spins $\bsigma = (\sigma_1, \ldots, \sigma_N)$, $\sigma_i \in \{-1,1\}$, arranged on a Bethe lattice with $p$-spin interactions, and interconnected with a one-dimensional chain. The Hamiltonian is given by
\begin{equation}
H(\bsigma) = -\sum_i \sigma_i h_i(\bsigma),
\end{equation}
with
\begin{eqnarray}
h_i(\bsigma) &=& \frac{1}{cp!}\sum_{j_1 \ldots j_{p-1}} J_{i,j_1,\ldots,j_{p-1}} c_{i,j_1,\ldots,j_{p-1}} \sigma_{j_1} \ldots \sigma_{j_{p-1}}\nonumber \\ 
&&+ \frac{J_0}{2}(\sigma_{i-1} + \sigma_{i+1}) \label{hfield} .
\end{eqnarray}
The connection strengths for the longe range $p$-spin interactions $J_{j_1 \ldots j_p}$ are taken from the distribution
\begin{equation}
P\left( J_{j_1 \ldots j_p} \right) = b\ \delta (J_{j_1 \ldots j_p} - J) + (1-b)\ \delta (J_{j_1 \ldots j_p} + J), \label{Jdist}
\end{equation}
with $b \in [0,1]$ the bias and $J$ a positive constant. The connection strength for the chain $J_0$ is a constant.
The couplings $c_{j_1 \ldots j_p} \in \{0,1\}$ are random, but with the restriction that every spin has degree $c$, or put otherwise, is part of $c$ hyperedges. These couplings are thus taken from the following distribution:
\begin{eqnarray}
P(c_{j_1 \ldots j_p}) &=& \left[ c\ \frac{(p-1)!}{N^{p-1}}\ \delta(c_{j_1 \ldots j_p} - 1) \right. \nonumber \\
 && \left. + \left(1 - c\  \frac{(p-1)!}{N^{p-1}}\right)\ \delta(c_{j_1 \ldots j_p}) \right] \nonumber \\
&&\times \prod_{j_1} \delta_{c, \sum_{j_2} \ldots \sum_{j_p} \frac{c_{j_1 \ldots j_p}}{(p-1)!}} .
\label{cdist}
\end{eqnarray}
The total number of hyperedges that a spin $\sigma$ is part of is then by definition $c$:
\begin{eqnarray}
\frac{1}{N}\sum_i \frac{1}{(p-1)!} \sum_{j_{1}} \ldots \sum_{j_{p-1}} c_{i,j_1, \ldots, j_{p-1}} &\equiv& c .
\end{eqnarray}
Since we are interested in the finitely connected case, we consider $c$ to be of order $\order(1)$, with $c/N \rightarrow 0$.

Self-interaction and hyperedges of reduced degree are prohibited by the following rule:
\begin{equation}
\forall k \neq l \in \{1, \ldots, p\}: j_k = j_l \Rightarrow c_{j_1 \ldots j_p} = 0 .
\end{equation}
This means that when any two indices are equal the hyperedge cannot exist. Also, the couplings are symmetrical: for any permutation $\pi$ in $\mathcal{S}_p$ we have that 
\begin{equation}
c_{j_1 \ldots j_p} = c_{j_{\pi(1)} \ldots j_{\pi(p)}}.
\end{equation}

\section{Saddle-point equations} \label{saddlesec}
As usual we replicate the partition function in order to calculate the free energy. Analytically, the calculation is analogous to the one found in \cite{hypergraphs}:
\begin{eqnarray}
f(\beta) &=& -\lim_{N\rightarrow \infty}\lim_{n\rightarrow 0}\frac{1}{\beta Nn} \log \Bra Z^n \Ket_{\bc, \bJ},
\end{eqnarray}
with $\Bra Z^n \Ket$ given, after some algebra, by the saddle point equation:
\begin{eqnarray}
\Bra Z^n \Ket &=& \frac{1}{\mathcal{N}} \int \{d\hat{P} dP\} \exp \left( N \Omega(P,\hat{P}) \right) \label{saddle}\\
\Omega(P,\hat{P}) &=& i \sum_{\bsigma} \hat{P}(\bsigma) P(\bsigma) \nonumber \\ 
&& + \frac{c}{p} \sum_{\btau_1 \ldots \btau_p} \prod_{k=1}^p P(\btau_k) \Bra e^{\frac{\beta J}{c} \sum_\alpha \tau_1^\alpha \ldots \tau_p^\alpha} \Ket_J -\frac{c}{p}  \nonumber \\
&&+ \frac{1}{N} \log \left[ \sum_{\bsigma_1 \ldots \bsigma_n} \exp \left( \beta J_0 \sum_{i \alpha} \sigma_i^\alpha \sigma_{i+1}^\alpha \right) \nonumber \right. \\ 
&&\left. \times \prod_i \left(-i \hat{P}(\bsigma_i)\right)^c\right] - \log(c!) \label{omega}
\end{eqnarray}
with $\alpha = 1, \ldots, n$ the replica index. The average over $J$ has to be taken according to equation (\ref{Jdist}), and represents the bond-disorder in the system. 

The next step is to take derivatives of $\Omega(P,\hat{P})$ to find the solution of our saddle point equation. For the sake of brevity we first define a new variable $F(\bsigma)$:
\begin{eqnarray}
F(\bsigma) &\equiv& c \sum_{\btau_1 \ldots \btau_{p-1}} \prod_{k=1}^{p-1} P(\btau_k) \Bra e^{\frac{\beta J}{c} \sum_\alpha \tau_1^\alpha \ldots \tau_{p-1}^\alpha \sigma^\alpha} \Ket_J \nonumber \\
\end{eqnarray}
with which we can write the solution of the saddle point equation as
\begin{eqnarray}
P(\bsigma) &=& \frac{\sum_{\bsigma^1 \ldots \bsigma^{n}}\left[\sum_i (F(\bsigma_i))^{-1} \delta_{\bsigma_i, \bsigma}\right] \prod_j T_{\bsigma_{j}, \bsigma_{j+1}}} {c^{-1}N \sum_{\bsigma^1 \ldots \bsigma^{n}} \prod_j T_{\bsigma_{j}, \bsigma_{j+1}}},  \nonumber \\ \label{saddle1} 
\end{eqnarray}
where we have introduced the transfer matrix
\begin{equation}
T_{\bsigma, \btau} \equiv \exp \left( \beta J_0 \sum_\alpha \sigma^\alpha \tau^\alpha \right) F^c(\bsigma). \label{transf}
\end{equation}
This equation can be simplified by introducing the left and right eigenvalue equations of this transfer matrix belonging to the largest eigenvalue $\lambda_0$:
\begin{eqnarray}
\sum_{\btau} T_{\bsigma, \btau} u(\btau) &=& \lambda_0 u(\bsigma) \label{eigright}\\
\sum_{\bsigma} v(\bsigma) T_{\bsigma, \btau} &=& \lambda_0 v(\btau) \label{eigleft}.
\end{eqnarray}
Inserting these properties in equation (\ref{saddle1}) leads to
\begin{eqnarray}
P(\bpsi) &=& \frac{\sum_{\bsigma \btau} T^{N-1}_{\bsigma \btau} Q_{\btau \bsigma}(\bpsi)}{\sum_{\bsigma} T^N{\bsigma \bsigma}} \\
&=& \frac{\sum_{\bsigma} u(\bsigma) v(\bpsi) \exp (\beta J_0 \sum_\alpha \sigma^\alpha \psi^\alpha) F^{c-1}(\bpsi)}{\lambda_0 \sum_{\bsigma} u(\bsigma) v(\bsigma)}, \nonumber \\ \label{Peq}
\end{eqnarray}
where $Q_{\bsigma, \btau} (\bpsi) \equiv \exp \left(\beta J_0 \sum_\alpha \sigma^\alpha \tau^\alpha \right) F^{c-1}(\bsigma) \delta_{\bsigma, \bpsi} $. In the transition to the last line we have used the fact that in the thermodynamic limit only this largest eigenvalue $\lambda_0$ contributes to the eigenvalue expansion of $T^N$. It is easy to show that this eigenvalue $\lambda_0 = 1$ in the limit $n \rightarrow 0$ (see e.g. \cite{reptrans}).

\section{RS analysis} \label{rs}
For the RS equations we assume that all the replicas of the system can be permuted without affecting the results. We make the usual assumption (see e.g. \cite{monasson}) that we can write the probability density $P(\bpsi)$ and the eigenvectors $u(\bpsi)$ and $v(\bpsi)$ as

\begin{eqnarray}
P(\bpsi) &=& \int dh\: W(h) \prod_{\alpha=1}^n \frac{e^{\beta h \psi^\alpha}}{2 \cosh (\beta h)} \label{FRS} \\
u(\bpsi) &=& \int dx\: \phi(x) \prod_{\alpha=1}^n e^{\beta x \psi^\alpha} \label{uRS} \\
v(\bpsi) &=& \int dy\: \chi(y) \prod_{\alpha=1}^n e^{\beta y \psi^\alpha}. \label{vRS}
\end{eqnarray}
Combining equations (\ref{uRS}) and (\ref{eigright}) allows us to calculate the density $\phi(x)$ in a self-consistent way. We also take the limit $n \rightarrow 0$, and obtain
\begin{eqnarray}
\phi(x') &=& \int dx\: \phi(x) \prod_{k=1}^{p-1} \prod_{\nu=1}^{c} \int dh_k^\nu W(h_k^\nu) \nonumber \\
&&\times \int dJ P(J) \delta\left[ x' - R(x, \bh)\right], \label{RS_phi}
\end{eqnarray}
with
\begin{eqnarray}
R(x, \bh) &\equiv& \frac{1}{\beta} \Bigg[
\sum_\nu \atanh \left(\tanh (\frac{\beta J}{c}) \prod_{k=1}^{p-1} \tanh (\beta h_k^\nu)\right) \nonumber \\
&&  + \atanh \left(  \tanh(\beta x) \tanh(\beta J_0) \right) \Bigg].
\end{eqnarray}
An analogous calculation leads to the self-consistent equation for $\chi(x)$:
\begin{eqnarray}
\chi(x') &=& \int dx\: \chi(x) \prod_{k=1}^{p-1} \prod_{\nu=1}^{c} \int dh_k^\nu W(h_k^\nu) \nonumber \\
&&\times \int dJ P(J) \delta\left[ x' - L(x, \bh)\right], \label{RS_chi}
\end{eqnarray}
where $L(x, \bh)$ is given by
\begin{eqnarray}
&&L(x, \bh) \equiv \frac{1}{\beta} \atanh \Bigg[\tanh(\beta J_0) \nonumber \\
&&\times \tanh \left( \beta x + \sum_\nu \atanh ( \tanh(\frac{\beta J}{c}) \prod_k \tanh(\beta h_k^\nu )) \right) \Bigg]. \nonumber \\
\end{eqnarray}
The third self-consistent equation is for the density $W(x)$. This can be found by using equation (\ref{Peq}) and filling in the RS ansatz for $P(\psi)$, yielding
\begin{eqnarray}
W(h) &=& \int dx dy\ \phi(x) \chi(y)\ \prod_{k=1}^{c-1} \prod_{l=1}^{p-1} dh_l^k W(h_l^k) dJ P(J) \nonumber \\
&&\times
\delta \left[ h - y - \frac{1}{\beta}\atanh(\tanh (\beta x) \tanh(\beta J_0)) \right. \nonumber \\
&&\left.  - \frac{1}{\beta}\sum_{k=1}^{c-1} \atanh\left(\tanh(\frac{\beta J}{c}) \prod_l \tanh(\beta h_l^k)\right) \right]. \nonumber \\ \label{RS_W}
\end{eqnarray}

The self-consistent equations above allow us to calculate the thermodynamic properties of the system. First of all, we find the magnetization as:
\begin{eqnarray}
m &=& \Bra \frac{1}{N} \sum_i \sigma_i \Ket_{\bc, \bJ} \\
&=& \int dh\: W(h) \tanh(\beta h). \label{mag}
\end{eqnarray}
Similarly we write down the EA-parameter $q_{\alpha\beta}$ (with $\alpha \neq \beta$):
\begin{eqnarray}
q_{\alpha\beta} &=& \Bra \left(\frac{1}{N} \sum_i \sigma_i^\alpha \sigma_i^\beta\right) \Ket_{\bc, \bJ} \\
&=& \int dh\: W(h) \tanh^2(\beta h). \label{ea}
\end{eqnarray}
The free energy is given by:
\begin{widetext}
\begin{eqnarray}
-\beta f(\beta) &=& \frac{c(1-p)}{p} \int dJ P(J)\left[ \prod_{k=1}^p 
\int dh_k\: W(h_k) \right]
\log \left( \sum_{\tau_1 \ldots \tau_p}e^{\beta \sum_k h_k \tau_k + \frac{\beta J}{c} \tau_1 \ldots \tau_p} \right)  \nonumber  \\
&&+  \int dJ P(J) \int dx\: \phi(x) \prod_{\nu=1}^c \prod_{k=1}^{p-1} \int dh_k^\nu W(h_k^\nu) \left( \frac{1}{2} \sum_s \log(G_s (x, \bh))
\right),
\end{eqnarray}
\end{widetext}

with
\begin{eqnarray}
G_{s} (x, \bh) &=& \left( \sum_{\tau} e^{\beta x \tau + \beta J_0 s \tau } \right) \nonumber \\
&&\times \prod_{\nu=1}^c \sum_{\tau_{1} \ldots \tau_{p-1}}  e^{\beta \sum_{k} h_k^\nu \tau_k + \frac{\beta J}{c} \tau_1 \ldots \tau_{p-1} s}.  \nonumber \\
\end{eqnarray}
By taking specific limits, these equations reduce to the self-consistent equations of other simpler systems. In the limit $J \rightarrow 0$ they reduce to those of an Ising chain. In the limits $\beta \rightarrow \infty$ and $J_0 \rightarrow 0$ they become the equations of the Bethe spin-glass at zero temperature (see e.g. \cite{bethezero}). In the latter case we notice that the distribution $W(h)$ corresponds to the cavity field distribution in the graph, and the distribution $\phi(x)$ corresponds to the effective local fields. The distribution $\chi(y)$ vanishes when $J_0 \rightarrow 0$.

\section{1RSB Self-consistent equations}\label{1rsb}
The RS assumption causes reentrance effects, indicating that replica symmetry is broken in certain regions of parameter space. As a first step to find improved results in this region, we can calculate the corresponding 1RSB equations describing the model. We start from equation (\ref{Peq}) and use the 1RSB form for the probability density $P(\bpsi)$ and the eigenvectors $u(\bpsi)$ and $v(\bpsi)$ (see e.g. \cite{monasson}):
\begin{eqnarray}
P(\bsigma) &=& \int \mathcal{D}W\ \Xi[W]\ \prod_{\gamma=1}^{\frac{n}{m}}\left[\int dhW(h)\frac{e^{\beta h\sum_{\alpha=1}^{m} \sigma_{\alpha,\gamma}}}
{[2\cosh(\beta h)]^m}\right] \nonumber \\
\label{eq:P_1RSB} \\
u(\bsigma)&=&\int \mathcal{D}\Phi\ \Omega[\Phi]\ \prod_{\gamma=1}^{\frac{n}{m}}\left[\int dx\Phi(x) e^{\beta x\sum_{\alpha=1}^{m} \sigma_{\alpha,\gamma}}\right]
\label{eq:L_1RSB} \\
v(\bsigma)&=&\int \mathcal{D}\Psi\ \Gamma[\Psi]\ \prod_{\gamma=1}^{\frac{n}{m}}\left[\int dy\Psi(y)e^{\beta y\sum_{\alpha=1}^{m} \sigma_{\alpha,\gamma}}\right].
\label{eq:R_1RSB} 
\end{eqnarray}
The objects $\Xi[W]$, $\Omega[\Phi]$ and $\Gamma[\Psi]$ are distributions of distributions, and the integrations $\int \mathcal{D} [\cdot]$ must be interpreted in a distributional sense. Replicas now acquire two indices, the first one $\gamma=1,\ldots,\frac{n}{m}$ denoting to which group of replicas it belongs and the second one $\alpha=1,\ldots,m$ indicating the particular replica in the group. The variable $m$ denotes the relative size of the group. We have to consider the limit $n\to 0$ and $m$ a real number, $m\in[0,1]$. This parameter $m$ is not a free parameter, we need to use the $m$ that extremizes the free energy (see e.g. \cite{mezpar}, \cite{janis}).

The detailed calculation of the self-consistent equations for these distributions can be found in the appendix (\ref{appendix1}). We now have three order parameters in the system: the magnetization $m$, and two Edwards-Anderson parameters, $q_0$ for replicas belonging to different pure states, and $q_1$ for replicas within a single pure state. They are given by

\begin{eqnarray}
m&=& \int \mathcal{D}W\ \Xi[W]\ \int dhW(h) \tanh (\beta h) \nonumber \\
q_0&=& \int \mathcal{D}W\ \Xi[W]\ \left[\int dhW(h) \tanh (\beta h)\right]^2 \nonumber \\
q_1&=& \int \mathcal{D}W\ \Xi[W]\ \int dhW(h) \left[\tanh (\beta h)\right]^2 \nonumber.
\end{eqnarray}
The free energy is given by
\begin{widetext}
\begin{eqnarray}
-\beta f(\beta) &=& \frac{c(1-p)}{m p}\prod_{k=1}^p \int \mathcal{D}W_k\ \Xi[W_k]\ \log \left[\int \frac{dh_k W_k(h_k)dJ P(J)}{[2\cosh(\beta h_k)]^m} \left( \sum_{\tau^1 \ldots \tau^p} \exp\left(\beta \sum_k h_k \tau^k + \frac{\beta J}{c}\tau^1 \ldots \tau^p \right) \right)^m \right] \nonumber \\
&&+ \frac{1}{m} \int \mathcal{D}\Phi\ \Omega[\Phi]\ \prod_{\nu=1}^c \prod_{k=1}^{p-1} 
\int \mathcal{D}W_{k, \nu}\ \Xi[W_{k,\nu}]\  \log \left(\mathcal{Z}_\Phi\left(\Phi,\{ W_{k,\nu}\}_{k,\nu} \right)\right), \label{freeE1rsb} 
\end{eqnarray}
\end{widetext}
with $\mathcal{Z}_\Phi$ a normalization
\begin{eqnarray}
\mathcal{Z}_\Phi\left(\Phi,\{ W_{k,\nu}\}_{k,\nu} \right) &\equiv& \int dx \Phi(x) \int \prod_{\nu,k} \frac{dh^{k,\nu} W_{k,\nu}(h^{k,\nu})}
{[2\cosh(\beta h^{k,\nu})]^m} \nonumber \\
&& \hspace{-3cm}\times \int dJ P(J) \exp \left( \frac{m}{2} \sum_\sigma \log G_s \left( x, \{h^{k,\nu}\}^{k,\nu} \right) \right).
\end{eqnarray}

\section{Numerical solution}\label{results}

The self-consistent equations above are numerically solved by the technique of population dynamics (\cite{mezpar}). Here the different probability densities that appear in equations (\ref{RS_phi}), (\ref{RS_chi}) and (\ref{RS_W}) are represented by distributions of fields, and by distributions of such distributions in the case of the corresponding 1RSB equations. A recursive scheme (see e.g. \cite{mezpar}) then allows these distributions to evolve towards the equilibrium distributions that solve the self-consistent equations. Monte-Carlo integration over these distributions allows to obtain the physical parameters. Furthermore, using the 1RSB equations, we have to determine the $m$ that extremizes $f(m)$. \\

The accuracy (and numerical cost) of the algorithm is mainly determined by the number of fields one chooses to make up the different distributions. This cannot be too low, since then the distributions will not be clearly outlined, but it cannot be too big either, since the numerical cost rises linearly (in the case of RS) and quadratically (in the case of 1RSB) with this parameter. In the RS algorithms we use populations of size $10^5$ fields, which results in smooth and accurate results. In the case of the 1RSB algorithm, however, we are constrained by the sheer numerical cost to populations of the order of $1000$ samples (so $1000$ populations of $1000$ fields each). This will put an upper limit to the accuracy we can achieve.

\subsection{RS results}\label{rsresults}

We test our results against those of \cite{bastian}. To this end, we put the system parameters to the following values: The degree of the interactions $p=2$, the number of connections per spin $c=6$, the connection strength $J=1$ and the bias $b = 0.625$. We are able to fully reproduce the RS results presented in \cite{bastian}, although we numerically solved the self-consistent equations (\ref{RS_phi}), (\ref{RS_chi}) and (\ref{RS_W}) to obtain these results, as opposed to the bifurcation analysis employed in \cite{bastian}.

\begin{figure}
\begin{center}
\includegraphics[width=.45\textwidth]{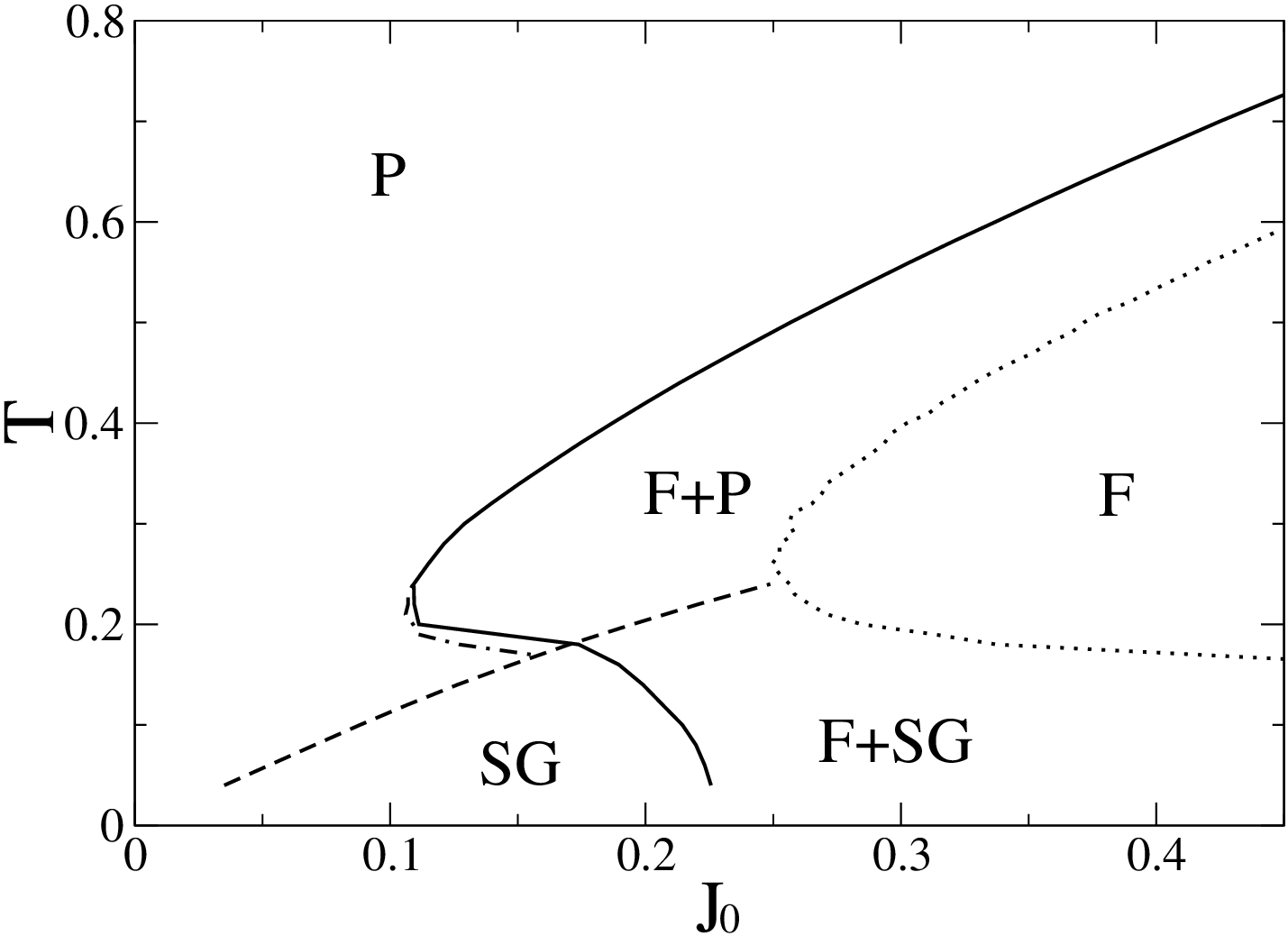}
\caption{The RS phase diagram for $p=3$, $c=3$, $J=1$ and $b=0.8$. The solid line depicts the ferromagnetic spinodal transition, and to the right of this line we can find ferromagnetic states, coexisting with the other solutions. The dotted line represents the ferromagnetic thermodynamic transition, where the ferromagnetic solution becomes thermodynamically stable. The spinglass transition line is given by the dashed line. The dash-dotted line indicates ergodicity breaking in the RS algorithm.} \label{phasediagram2}
\end{center}
\end{figure}

We are interested in the phase diagrams for systems with bias and hyperspin interactions ($p>2$). A typical result can be found in Fig. \ref{phasediagram2}, where we plot the full phase diagram for the parameters $p=3$, $c=3$, $J=1$ and $b=0.8$. The spinodal transition, in full, is of first order, which is typical for hyperspin interactions. This also means we will have a coexistence of solutions close to this transition. By comparing the corresponding free energies we determine the thermodynamic transition, given as a dotted line. The different phases and coexisting phases are indicated in Fig. \ref{phasediagram2}.

Furthermore, we clearly see that when the ferromagnetic transition lines approach the spinglass transition, a reentrance effect sets in, indicative of replica symmetry breaking. By using different initial distributions in the RS algorithm and comparing the final distributions for quenched disorder, we check whether ergodicity is broken or not in the reentrance region. These points are indicated in Fig. \ref{phasediagram2} as a dash-dotted line that further illustrates the broken replica symmetry.

\begin{figure}
\begin{center}
\includegraphics[width=.45\textwidth]{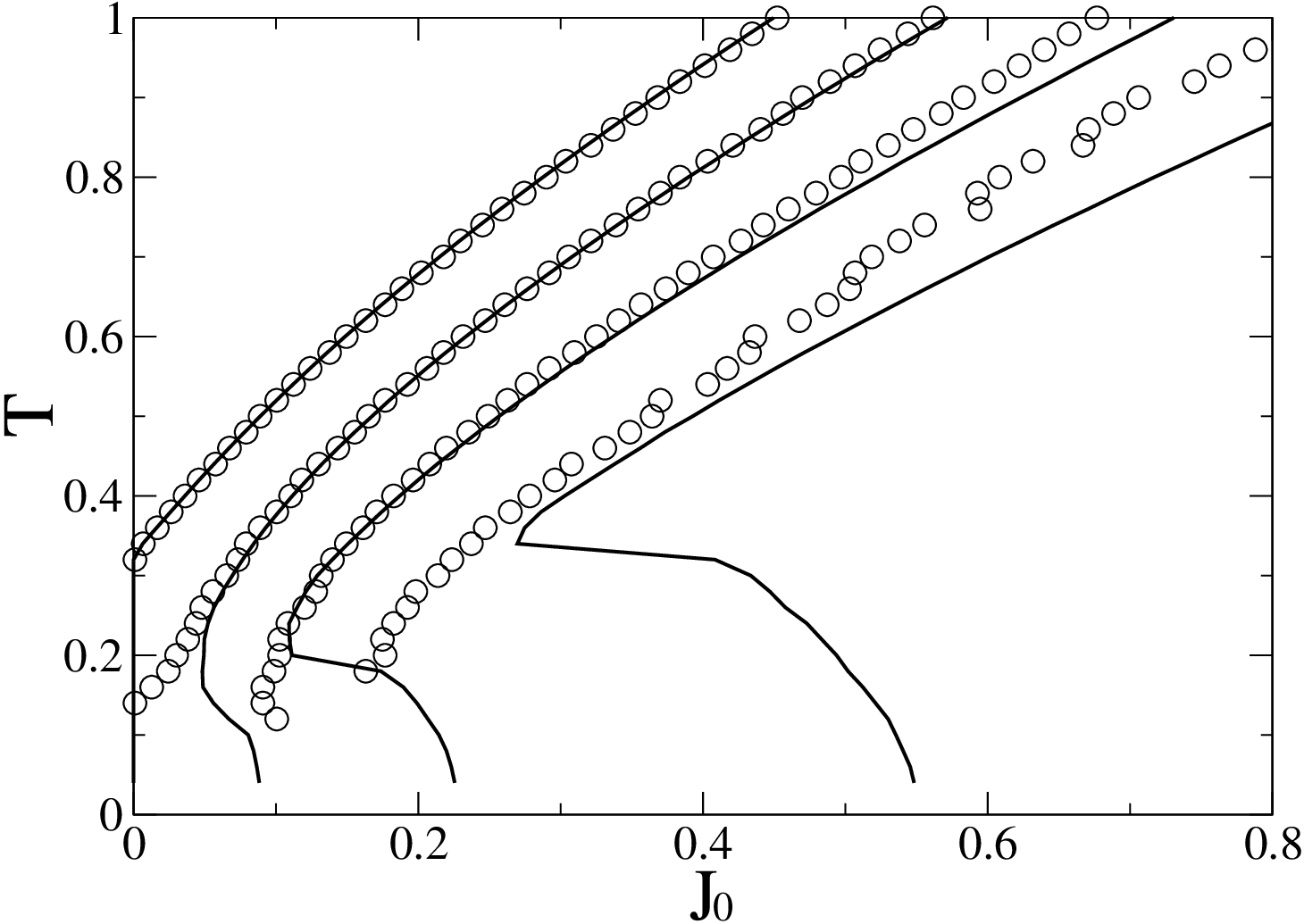}
\caption{The RS phase diagram for $p=3$, $c=3$, $J=1$ and different biases $b$. The solid lines represent the dynamic ferromagnetic transition, with from left to right: $b=1$, $b=0.9$, $b=0.8$ and $b=0.7$. The ferromagnetic region is located to the right of the respective transition lines. The circles are the corresponding transitions obtained by simulations. See Fig. \ref{phasediagram2} for a further specification of the different phases.} \label{phasediagram1}
\end{center}
\end{figure}

Next, to check the influence of frustration on these results, we can generate similar phase diagrams for other values of the bias $b$. The spinodal transitions are given in Fig. \ref{phasediagram1}, with $p=3$, $c=3$ and several values of the bias $b$, along with Monte-Carlo simulation results. Again we find reentrance effects for the frustrated graphs, and this effect becomes stronger when the frustration increases due to smaller bias $b$. The simulations agree with the results obtained by population dynamics for low frustration, but start to differ slightly when this frustration increases due to the slow dynamics of the system for these parameters. 

\subsection{1RSB results}\label{1rsbresults}
Since the above results all display reentrance indicating RSB, we now look closer to these regions in parameter space employing the 1RSB approach to determine possible corrections. However, the algorithm proves to be very demanding numerically, and hence the results turn out to be rather noisy. The free energy results presented in this section are averages over many iterations of the algorithm.

\begin{figure}
\vspace{0.5cm}
\begin{center}
\includegraphics[width=.45\textwidth]{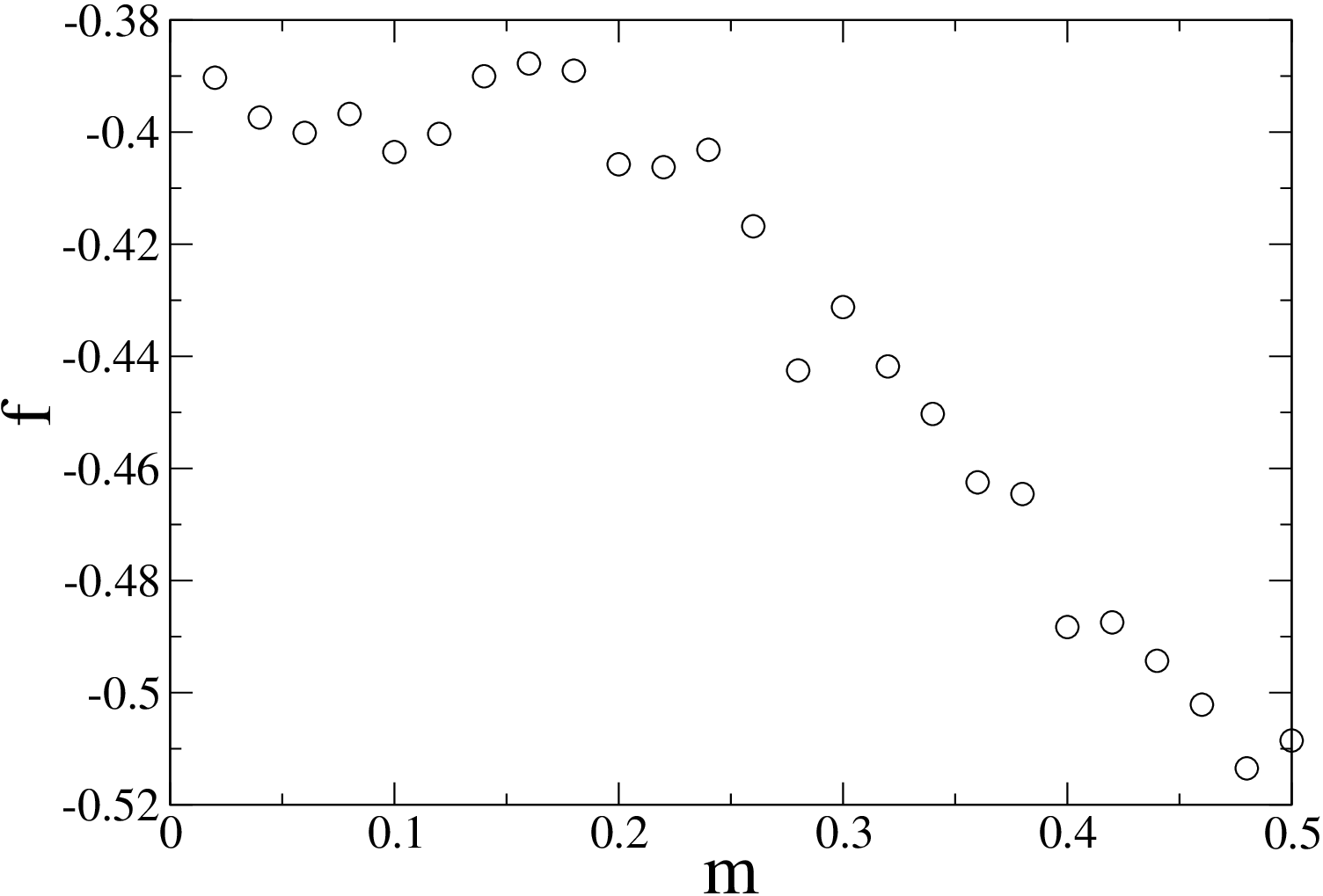}
\caption{The 1RSb free energy as  function of $m$ for  $T=0.15$, $J_0=0.15$, $p=3$, $c=3$, $J=1$ and $b=0.8$. The maximum is reached around $m_{\text{max}} \approx 0.15$} \label{rsbfig1}
\end{center}
\end{figure}

\begin{figure}
\vspace{0.5cm}
\begin{center}
\includegraphics[width=.45\textwidth]{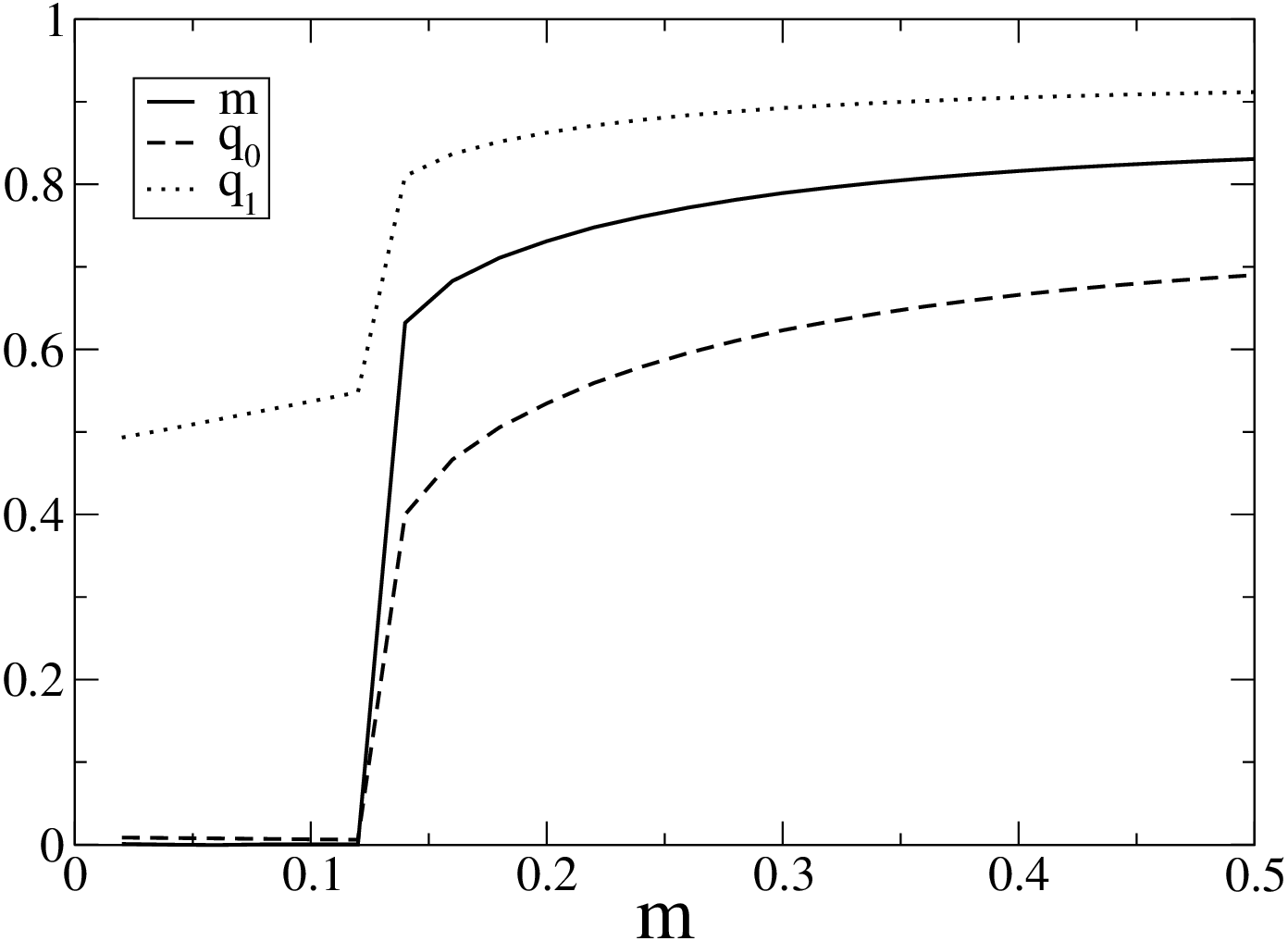}
\caption{The 1RSb order parameters as  function of $m$ for  $T=0.15$, $J_0=0.15$, $p=3$, $c=3$, $J=1$ and $b=0.8$. The full line is the magnetization, the dashed line is $q_0$, and the dotted line is $q_1$. At $m_{\text{max}}$ (see Fig. 3) the magnetization starts to be nonzero, indicating a ferromagnetic phase.} \label{rsbfig2}
\end{center}
\end{figure}

In order to see whether the 1RSB approach will give different results in the reentrance region, we focus on a single point in this region on Fig. \ref{phasediagram2}: the point $T=0.15$ and $J_0=0.15$. The free energy as a function of the RSB parameter $m$ for this point is plotted in Fig. \ref{rsbfig1}. We need to first find that value of $m$ that maximizes this free energy, and although the results are noisy we  see that this maximum is reached around $m_{\text{max}} \approx 0.15$. The order parameters of the system are plotted as function of $m$ in Fig. \ref{rsbfig2}, and we see that at $m_{\text{max}}$ the magnetization is nonzero, indicating that this point is located inside the ferromagnetic region, as opposed to the RS spinglass result. However, because of the noise in the system and the sharp transition from a spinglass solution to a ferromagnetic solution in the neighborhood of $m_{\text{max}}$, it is hard to present quantitative results with the 1RSB approach. We did find a qualitative indication that the 1RSB method will at least decrease the reentrance region in the phase diagram. Creating a 1RSB phase transition line between the ferromagnetic and spinglass phase is unfortunately beyond our present computational capabilities, and we will not pursue this goal in this paper.

We have also calculated the limits of the 1RSB equations for $T \rightarrow 0$ and $J_0 \rightarrow 0$. The results obtained are in agreement with those presented in \cite{bethezero}. The noise in the system seems to decrease significantly in this limiting case, and we  conclude that the major source of the noise in the algorithm is the interaction between the ferromagnetic ring and the bond-disordered random hypergraph. When the ring is not present, we are left with only a single density, and the algorithm functions much better for the relatively small system sizes we employ.

\section{Conclusions}\label{discussion}
In this paper, we have studied the thermodynamic properties of a Bethe lattice with p-spin interactions augmented with a one-dimensional ferromagnetic chain interconnecting all spins. We have calculated the RS and 1RSB self-consistent equations describing the order parameters and the free energy. By employing population dynamics we have been able to solve these self-consistent equations and to obtain the phase diagrams of the system as function of the different system parameters. We have noticed reentrance effects in the RS phase diagrams, increasing with the bond-disorder in the system, indicating RSB. The 1RSB algorithm indicates that this reentrance region will decrease when replica symmetry breaking is taken into account. The algorithm is quite noisy however, preventing us from giving precise quantitative results. By both the analytic analysis and the numerical results using the corresponding algorithms at temperature zero and zero chain interaction, we  observe a  substantial decrease in the noise on the results. This leads us to believe that the ferromagnetic chain is the major factor causing the noise in the 1RSB algorithm. 

\begin{acknowledgments}
We would like to thank N. S. Skantzos for interesting and informative discussions about the results obtained.
\end{acknowledgments}

\begin{widetext}
\appendix 
\section{Calculation of the self-consistent equations} \label{appendix1}
To calculate the 1RSB self-consistent equations for the distributions $\Xi[W]$, $\Omega[\Phi]$ and $\Gamma[\Psi]$, we start with the eigenvector equation of the transfer matrix $T$ (\ref{eigright}), and fill in the 1RSB assumption concerning the different matrices and vectors (\ref{eq:P_1RSB}) and (\ref{eq:L_1RSB}):
\begin{eqnarray}
\sum_{\btau} T_{\bsigma, \btau} u(\btau) &=& \sum_{\btau} \exp \left( \beta J_0 \sum_{\alpha=1}^n \sigma_\alpha \tau_\alpha \right) \left[ \sum_{\btau^1 \ldots \btau^{p-1}} \prod_{k=1}^{p-1} P(\btau^k) \left( e^{\frac{\beta J}{c} \sum_\alpha \tau^1_\alpha \ldots \tau^{p-1}_\alpha \sigma_\alpha} \right)
\right]^c u(\btau) \nonumber \\
&=& \sum_{\btau} \int \mathcal{D}\Phi\ \Omega[\Phi]\ \prod_{\gamma=1}^{\frac{n}{m}} \int dx_\gamma \Phi(x_\gamma) \exp \left(\beta x_\gamma \sum_{\alpha=1}^{m} \tau_{\alpha,\gamma} + \beta J_0 \sum_\gamma \sum_\alpha \sigma_{\alpha,\gamma} \tau_{\alpha,\gamma}\right) \nonumber \\
&&\times
\prod_{\nu=1}^c \sum_{\btau^{1,\nu} \ldots \btau^{p-1,\nu}} \prod_{k=1}^{p-1} 
\int \mathcal{D}W_{k, \nu}\ \Xi[W_{k,\nu}]\ \prod_{\gamma=1}^{\frac{n}{m}}\int dh^{k,\nu}_\gamma W_{k,\nu}(h^{k,\nu}_\gamma)\frac{e^{\beta h^{k,\nu}_\gamma \sum_{\alpha=1}^{m} \tau^{k,\nu}_{\alpha,\gamma}}}
{[2\cosh(\beta h^{k,\nu}_\gamma)]^m} \nonumber \\
&&\times 
\exp \left(\frac{\beta J}{c} \sum_\alpha \sum_\gamma \tau^{1,\nu}_{\alpha,\gamma} \ldots \tau^{p-1}_{\alpha,\gamma} \sigma_{\alpha,\gamma}\right) .
\label{blabla}
\end{eqnarray}
We put all the exponentials together 
\begin{eqnarray}
F_\gamma^{\sigma_{\alpha,\gamma}} \left( x_\gamma, \{h^{k,\nu}_\gamma\}^{k,\nu} \right) &\equiv&
\left( \sum_{\tau} \exp \left(\beta x_\gamma \tau + \beta J_0 \sigma_{\alpha,\gamma} \tau \right) \right) \nonumber \\
&&\times
\prod_{\nu=1}^c \sum_{\tau^{1,\nu} \ldots \tau^{p-1,\nu}} \prod_{k=1}^{p-1} 
\exp \left(\beta h^{k,\nu}_\gamma \tau^{k,\nu} + \frac{\beta J}{c} \tau^{1,\nu} \ldots \tau^{p-1,\nu} \sigma_{\alpha,\gamma}\right), \label{F_rsb}
\end{eqnarray}
and use the identity
\begin{eqnarray}
\prod_{\gamma,\alpha} F_\gamma^{\sigma_{\alpha,\gamma}} &=& \exp \left( \sum_{\gamma,\alpha} \log F_\gamma^{\sigma_{\alpha,\gamma}} \right) \nonumber \\
&=& \exp \left( \sum_{\gamma,\alpha}  \sigma_{\alpha,\gamma} \left( \frac{1}{2} \sum_\sigma \sigma  \log F_\gamma^{\sigma}\right) + m \sum_\gamma \left( \frac{1}{2} \sum_\sigma \log F_\gamma^{\sigma}\right) \right) \label{trick1}
\end{eqnarray}
to obtain
\begin{eqnarray}
\sum_{\btau} T_{\bsigma, \btau} u(\btau) 
&=& 
\int \mathcal{D}\Phi\ \Omega[\Phi]\ \prod_{\gamma=1}^{\frac{n}{m}} \int dx_\gamma \Phi(x_\gamma)  
\prod_{\nu=1}^c \prod_{k=1}^{p-1} 
\int \mathcal{D}W_{k, \nu}\ \Xi[W_{k,\nu}]\ \prod_{\gamma=1}^{\frac{n}{m}}\int \frac{dh^{k,\nu}_\gamma W_{k,\nu}(h^{k,\nu}_\gamma)}
{[2\cosh(\beta h^{k,\nu}_\gamma)]^m} \nonumber \\
&&\times 
\prod_\gamma \exp \left( \sum_{\alpha} \sigma_{\alpha,\gamma} \left( \frac{1}{2} \sum_\sigma \sigma  \log F_\gamma^{\sigma}\left( x_\gamma, \{h^{k,\nu}_\gamma\}^{k,\nu} \right)\right) + m \left( \frac{1}{2} \sum_\sigma \log F_\gamma^{\sigma}\left( x_\gamma, \{h^{k,\nu}_\gamma\}^{k,\nu} \right)\right) \right).\nonumber \\
\end{eqnarray}
We now employ the fact that 
\begin{eqnarray}
\frac{1}{2\beta} \sum_\sigma \sigma  \log F_\gamma^{\sigma}\left( x_\gamma, \{h^{k,\nu}_\gamma\}^{k,\nu} \right) &\equiv& R(x^\gamma, \{h^{k,\nu}_\gamma\}^{k,\nu}) \\
&=& \frac{1}{\beta} \left[
\sum_\nu \atanh \left(\tanh (\frac{\beta J}{c}) \prod_{k=1}^{p-1} \tanh (\beta h_\gamma^{k,\nu})\right) + \atanh \left(  \tanh(\beta x^\gamma) \tanh(\beta J_0) \right) \right], \nonumber
\end{eqnarray}
and write the expression (\ref{blabla}) as a functional integral over a delta function. This leads to 
\begin{eqnarray}
\sum_{\btau} T_{\bsigma, \btau} u(\btau) 
&=&
\int \mathcal{D}\Phi'\
\Bigg[ \int \mathcal{D}\Phi\ \Omega[\Phi]\ \prod_{\nu=1}^c \prod_{k=1}^{p-1} 
\int \mathcal{D}W_{k, \nu}\ \Xi[W_{k,\nu}]\  \nonumber \\
&& \times \delta_F\Bigg( \Phi'(x') - \int \frac{dx \Phi(x)}{\mathcal{Z}_\Phi\left(\Phi,\{ W_{k,\nu}\}_{k,\nu} \right)}  \int \prod_{\nu=1}^c \prod_{k=1}^{p-1} \frac{dh^{k,\nu} W_{k,\nu}(h^{k,\nu})}
{[2\cosh(\beta h^{k,\nu})]^m} \exp \left( \frac{m}{2} \sum_\sigma \log F^{\sigma}\left( x, \{h^{k,\nu}\}^{k,\nu} \right) \right)  \nonumber \\ 
&&\times
\delta\left(x' - R(x^\gamma, \{h^{k,\nu}_\gamma\}^{k,\nu})\right) \Bigg)\Bigg]\left(\mathcal{Z}_\Phi\left(\Phi,\{ W_{k,\nu}\}_{k,\nu} \right)\right)^{\frac{n}{m}}\prod_{\gamma=1}^{\frac{n}{m}}\int dz \Phi'(z) \exp \left( \beta z\sum_{\alpha} \sigma_{\alpha,\gamma} \right).
\end{eqnarray}
The functional $\mathcal{Z}_\Phi$ is a normalization factor given by
\begin{eqnarray}
\mathcal{Z}_\Phi\left(\Phi,\{ W_{k,\nu}\}_{k,\nu} \right) \equiv \int dx \Phi(x) \int \prod_{\nu=1}^c \prod_{k=1}^{p-1} \frac{dh^{k,\nu} W_{k,\nu}(h^{k,\nu})}
{[2\cosh(\beta h^{k,\nu})]^m} \exp \left( \frac{m}{2} \sum_\sigma \log F^{\sigma}\left( x, \{h^{k,\nu}\}^{k,\nu} \right) \right).
\end{eqnarray}
We now identify this with (\ref{eq:R_1RSB}) to find the self-consistent equation for the functional distribution $\Omega[\Phi]$. We also take the limit $n \rightarrow 0$.
\begin{eqnarray}
\Omega[\Phi'] &=&
\int \mathcal{D}\Phi\ \Omega[\Phi]\ \prod_{\nu=1}^c \prod_{k=1}^{p-1} 
\int \mathcal{D}W_{k, \nu}\ \Xi[W_{k,\nu}]\  \\
&& \times \delta_F\Bigg( \Phi'(x') - \int \frac{dx \Phi(x)}{\mathcal{Z}_\Phi\left(\Phi,\{ W_{k,\nu}\}_{k,\nu} \right)}  \int \prod_{\nu=1}^c \prod_{k=1}^{p-1} \frac{dh^{k,\nu} W_{k,\nu}(h^{k,\nu})}
{[2\cosh(\beta h^{k,\nu})]^m} \exp \left( \frac{m}{2} \sum_\sigma \log F^{\sigma}\left( x, \{h^{k,\nu}\}^{k,\nu} \right) \right)  \\ 
&&\times
\delta\left(x' - R(x^\gamma, \{h^{k,\nu}_\gamma\}^{k,\nu})\right) \Bigg) \label{OmegaPhi}
\end{eqnarray}
A similar calculation yields analogous equations for the two other distributions $\Omega[\Phi]$ and $\Gamma[\Psi]$.
\end{widetext}

\end{document}